\documentclass{article}
\usepackage[preprint]{spconf}
\usepackage{spconf,amsmath,graphicx,hyperref}

\usepackage{tabularx}
\usepackage{makecell}
\usepackage{multirow}
\usepackage{booktabs}
\usepackage{algorithmic}
\usepackage{amsfonts}
\usepackage{threeparttable}
\usepackage{xcolor}
\usepackage{float}
\usepackage{textgreek}

\toappear{
  \raisebox{-5mm}{
    \parbox[b]{\textwidth}{\footnotesize\raggedright
© 2026 IEEE. Personal use of this material is permitted. Permission from IEEE must be obtained for all other uses, in any current or future media, including reprinting/republishing this material for advertising or promotional purposes, creating new collective works, for resale or redistribution to servers or lists, or reuse of any copyrighted component of this work in other works.
}}}

\title{Time vs. Layer: Locating Predictive Cues for Dysarthric Speech Descriptors in wav2vec 2.0}
\name{Natalie Engert, Dominik Wagner, Korbinian Riedhammer, Tobias Bocklet}
\address{Technische Hochschule Nürnberg Georg Simon Ohm, Germany}% \\\texttt{firstname.lastname@th-nuernberg.de}}
\begin{document}

\ninept
\maketitle
\begin{abstract}

Wav2vec 2.0 (W2V2) has shown strong performance in pathological speech analysis by effectively capturing the characteristics of atypical speech. 
Despite its success, it remains unclear which components of its learned representations are most informative for specific downstream tasks.
In this study, we address this question by investigating the regression of dysarthric speech descriptors using annotations from the Speech Accessibility Project dataset. We focus on five descriptors, each addressing a different aspect of speech or voice production: intelligibility, imprecise consonants, inappropriate silences, harsh voice and monoloudness. Speech representations are derived from a W2V2-based feature extractor, and we systematically compare layer-wise and time-wise aggregation strategies using attentive statistics pooling. Our results show that intelligibility is best captured through layer-wise representations, whereas imprecise consonants, harsh voice and monoloudness benefit from time-wise modeling. For inappropriate silences, no clear advantage could be observed for either approach.

\end{abstract}
\section{Introduction}
\label{sec:intro}
Pretrained wav2vec 2.0 (W2V2)~\cite{baevski2020wav2vec} models have become a popular backbone architecture for various pathological speech processing tasks~\cite{hu_dysarthric_asr_2023,w2v2_robustness, stuttering_paper,wagner23_interspeech}.
Beyond linguistic and speaker-related features, W2V2 embeddings capture paralinguistic properties such as voice quality, prosody and speaking style, 
making them well-suited for 
clinical applications including speech quality assessment~\cite{speech_quality_w2v2, children_speech}, phoneme-level analysis~\cite{baumann2024towards}, and adaptation of automatic speech recognition systems for dysarthric speech~\cite{speaker_adaption_w2v2}.
A common approach is to select the transformer layer providing the most relevant representations and apply temporal aggregation (e.g., mean pooling) to obtain fixed-size utterance-level features for downstream tasks~\cite{javanmardi2023wav2vec,narain25_interspeech,speaker_verification_w2v2,wagner24b_interspeech,wagner25_interspeech}.
Despite the widespread use of this methodology, limited understanding regarding the specific information captured by individual transformer layers remains, making the selection of appropriate layers a non-trivial challenge~\cite{layer_problem_pathological_speech}.

Pasad et al.\ \cite{layerwise_analysis_asru} demonstrate that
% In~\cite{layerwise_analysis_asru}, 
earlier layers of W2V2 were associated with local acoustic features, while intermediate layers encoded phoneme identity, word identity and meaning. 
% As the experiments were conducted on the LibriSpeech~\cite{librispeech} dataset, which consists of typical, well-articulated speech, the transferability of these findings to quality assessment of pathological speech remains unclear. 
Because the experiments were conducted on the LibriSpeech \cite{librispeech} dataset, which contains clear, well-articulated speech, it is uncertain how far these results generalize to the quality assessment of pathological speech.
 
Nguyen et al.\ \cite{nguyen_asr_slt2024} conducted a layer-wise analysis of large W2V2-based models and demonstrated that intermediate layers offered the most informative features for rating severity and intelligibility in head and neck cancer speech.
Similarly, Wiepert et al.\ \cite{layer_problem_pathological_speech} reported optimal classification of pathological speech traits (e.g., strained voice, irregular articulatory breakdowns, rapid speech rate) using earlier to intermediate layers.
In a subsequent layer-wise study, Javanmardi et al.\ \cite{javanmardi2023wav2vec} utilized
%Another layer-wise analysis was conducted in~\cite{javanmardi2023wav2vec}, using 
a W2V2 base model to classify both dysarthria and its severity levels on the UASpeech dataset~\cite{uaspeech_database}.
Consistent with previous findings, the study reported that earlier layers were particularly effective for distinguishing dysarthric from healthy speech, while later layers were more suitable for classifying severity levels.  
Together, these studies indicate that although certain individual layers often show particular suitability for a given task, these findings are not necessarily generalizable across different datasets or clinical speech assessment tasks.

Furthermore, statistical pooling compresses temporal information into global summary statistics, which may obscure clinically relevant temporal cues.
For instance, studies by Baumann et al.\ \cite{baumann2024towards,baumann24b_interspeech} demonstrate that classifying speech patterns such as hypernasality and articulatory tension in cleft lip and palate speech often depends on specific phonemes. Their findings
suggest that classifier decisions may be influenced by temporally localized information, highlighting the importance of preserving temporal structure for pathological speech analysis.

In this paper, we investigate the impact of different attentive pooling strategies on the regression of dysarthric speech descriptors by comparing attentive pooling over time with attentive pooling applied across transformer layers. We focus on W2V2 in our investigation, as it is among the most widely used feature extractors in pathological speech processing.
Our contributions are:
\begin{itemize}
  \item A comparative evaluation of time-wise vs.\ layer-wise attentive pooling mechanisms for dysarthric speech descriptor regression.
  \item An analysis of the impact of attention head count on model performance.
  \item A visual exploration of attention patterns across transformer layers.
\end{itemize}

To the best of our knowledge, this work presents the first systematic analysis of automatic speech quality assessment for these specific descriptors, offering insights into how time-wise and layer-wise feature information predicts clinically relevant speech characteristics.

\section{Data}
The Speech Accessibility Project~\cite{sap_paper} (SAP) dataset aims to facilitate research and technological development in dysarthric speech recognition. 
We use the 2024-11-30 release, containing recordings from 430 participants with Parkinson’s disease (PD), amyotrophic lateral sclerosis (ALS), cerebral palsy, stroke, or Down syndrome. These conditions are commonly associated with dysarthria, a motor speech disorder affecting articulation, prosody and resonance~\cite{sap_cat_artikel}.

A subset of the recordings 
%in the SAP 2024-11-30 dataset 
has been annotated using the Mayo Clinic rating system~\cite{sap_cat_artikel}, a widely adopted framework for characterizing and categorizing neurological speech disorders. Each speech descriptor is rated on a 7-point rating scale (1=typical, 7=severe).
%, where a score of 1 corresponds to typical speech production, and a score of 7 indicates a severe deviation from expected speech characteristics. 
Annotations comprise up to 12 of 45 available speech descriptors per labeled sample, although the number of annotated descriptors varies across the dataset. 
%and only a subset includes labels for all 12 descriptors. 
For consistency, we focus on five descriptors that capture both global and local speech characteristics: intelligibility, imprecise consonants, inappropriate silences, harsh voice, and monoloudness.
\begin{table}[tb]
\setlength{\tabcolsep}{5pt}
\caption{Number of voice recording samples (Smpl) and speakers (Spk) for each subset corresponding to the selected speech descriptors.}
% Each speech descriptor is associated with its own dataset.
\centering
\resizebox{\columnwidth}{!}{%
{\scriptsize
\begin{tabular}{c|cc|cc|cc}
\toprule
\multirow{2}{*}{\textbf{Subsets}} & \multicolumn{2}{c}{\textbf{Train}} & \multicolumn{2}{c}{\textbf{Dev}} & \multicolumn{2}{c}{\textbf{Test}} \\
 & Smpl & Spk & Smpl & Spk & Smpl & Spk \\
\midrule
Intelligibility        & 5644 & 208 & 916 & 32 & 912 & 32 \\
Imprecise consonants   & 5647 & 209 & 914 & 32 & 909 & 32 \\
Inappropriate silences & 5137 & 182 & 900 & 32 & 910 & 32 \\
Harsh voice            & 5650 & 209 & 916 & 32 & 911 & 32 \\
Monoloudness           & 5650 & 209 & 916 & 32 & 911 & 32 \\
\bottomrule
\end{tabular}} }
\label{tab:dataset_statistics}
\end{table}

To maximize training data, 
%for the models, separate subsets were created for each of the five selected speech descriptors. 
descriptor-specific subsets were created, including all samples rated for the respective descriptor.
%Each descriptor-specific subset comprises all samples with available ratings for the respective descriptor. 
Speaker-exclusive splits were applied, with the same test speakers used across all subsets to ensure fair and comparable evaluation across descriptors.
The dataset statistics are presented in Table \ref{tab:dataset_statistics}.
In our subsets, recordings are dominated by Parkinson’s disease ($\sim$80-90\%), with ALS ($\sim$8-10\%), cerebral palsy ($\sim$7\%), and Down syndrome ($\sim$3-4\%) less represented. 
%Inappropriate silences are rarely annotated for cerebral palsy and Down syndrome ($\sim$1\% each).

\section{Methods} 

\begin{figure}[h]
\centerline{\includegraphics[width=0.93\linewidth]{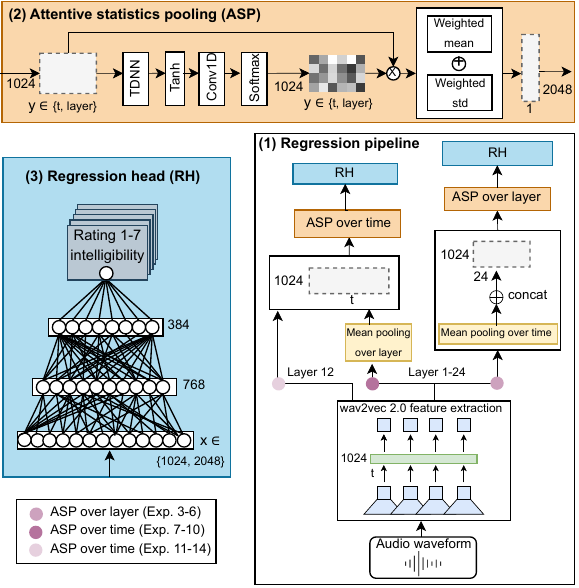}}
\caption{Overview of all experimental attention configurations in our regression pipeline. %(1).
%Component (2) visualizes the channel- and context-dependent attentive statistics pooling mechanism, which is introduced in~\cite{ecapa_tdnn} and employed as the layer-wise and time-wise attentive pooling method in our experiments. Component (3) illustrates the regression head used to predict the speech descriptor ratings.
}
\label{fig_overview_architecture}
\end{figure}

% This section outlines the methodology employed for the regression of speech descriptors annotated in the SAP dataset. 
For the regression tasks, we utilize a W2V2-based feature extractor, followed by either a layer-wise or a time-wise attentive pooling mechanism. 
The extracted features are subsequently fed into a regression head, implemented as a standard fully connected feedforward neural network with a ReLU activation function in the hidden layers. The output layer consists of a single neuron, producing a continuous value as the final regression output. Regression was employed rather than classification, as it naturally handles imbalanced rating distributions, preserves the ordinal structure of the scores, and reflects their continuous nature.
Figure \ref{fig_overview_architecture} provides an overview of the components used for our experiments. 

\subsection{Feature Extraction: Wav2vec 2.0}
W2V2 is a transformer-based model designed to learn speech representations directly from raw audio \cite{baevski2020wav2vec}. Pretrained on large unlabeled speech corpora, it captures diverse acoustic patterns. The large configuration (311M parameters) includes a convolutional encoder that converts waveforms into latent features, followed by 24 transformer blocks using self-attention to contextualize these features temporally.
% W2V2 is a transformer-based architecture designed to learn meaningful speech representations directly from raw audio waveforms\cite{baevski2020wav2vec}.
The W2V2 features used for our experiments were extracted from a W2V2-large-XLSR model\footnote{https://huggingface.co/jonatasgrosman/wav2vec2-large-xlsr-53-english} which is pretrained on more than 50k hours of speech in 53 languages and fine-tuned for speech recognition in English on the Mozilla Common Voice 6.1 dataset~\cite{common_voice}. It produces 1024-dimensional embeddings approximately every 20 milliseconds of audio input, with a receptive field of 25 milliseconds.
% To ensure consistency across experiments and to isolate the effect of embedding aggregation strategies, the weights of the feature extractor are not updated during training.
To maintain consistency and isolate embedding aggregation effects, the feature extractor weights remain frozen during training.

\renewcommand{\arraystretch}{0.8} % Standard ist 1.0

\begin{table*}[t]
\centering
\caption{
Overview of layer-wise and time-wise ASP results. The models were evaluated using the Pearson correlation coefficient (PCC, $\uparrow$) and mean squared error (MSE, $\downarrow$). PCC and MSE values are reported for each speech descriptor on the test split. 
%Average values were computed separately for both attention groups.
}
\resizebox{\textwidth}{!}{%
{\tiny
\begin{tabular}{c|c|c|c||cc||cc||cc||cc||cc}
\toprule
\multirow{2}{*}{\textbf{Exp.}} & \multirow{2}{*}{\textbf{Layer}} & \multirow{2}{*}{\textbf{Time}} & \multirow{2}{*}{\makecell{\textbf{Att.} \\ \textbf{Heads}}} &
\multicolumn{2}{c||}{\textbf{Intelligibility}} &
\multicolumn{2}{c||}{\textbf{Impr. consonants}} &
\multicolumn{2}{c||}{\textbf{Inappr. silences}} &
\multicolumn{2}{c||}{\textbf{Harsh voice}} &
\multicolumn{2}{c}{\textbf{Monoloudness}} \\
& & & & PCC & MSE & PCC & MSE & PCC & MSE & PCC & MSE & PCC & MSE \\
%%%%%%%%%%%%%%%%% Baseline
\midrule
1 & Mean & Mean & -   & 0.684 & 0.760 & 0.788 & 0.440 & 0.688 & 0.228 & 0.636 & 0.929 & 0.551 & 0.866 \\
2 & 12   & Mean & -   & 0.690 & 0.764 & 0.783 & 0.437 & 0.706 & 0.223 & 0.574 & 1.059 & 0.558 & 0.859 \\
\midrule
%%%%%%%%%%%%%%%%%%%% ASP Layer
\multicolumn{14}{c}{\textbf{ASP OVER LAYER}} \\
\midrule
3  & ASP & Mean & 1   & 0.650 & 0.747 & 0.778 & 0.449 & 0.692 & 0.219 & 0.648 & 0.902 & 0.563 & 0.835 \\
4  & ASP & Mean & 5   & \textbf{0.696} & 0.725 & 0.793 & 0.428 & 0.707 & 0.220 & 0.624 & 0.959 & 0.554 & 0.856 \\
5  & ASP & Mean & 64  & 0.673 & 0.724 & 0.783 & 0.436 & 0.698 & \textbf{0.214} & 0.631 & 0.949 & 0.554 & 0.849 \\
6 & ASP & Mean & 128 & 0.688 & \textbf{0.723} & 0.791 & 0.420 & 0.697 & 0.225 & 0.626 & 0.958 & 0.567 & 0.831 \\
\cmidrule{1-14}
%\multicolumn{4}{c||}{Average} & 0.677 & 0.730 & 0.786 & 0.433 & 0.698 & 0.220 & 0.632 & 0.942 & 0.559 & 0.843 \\
%\cmidrule{1-14}

%%%%%%%%%%%%%%%% ASP over Time
\multicolumn{14}{c}{\textbf{ASP OVER TIME}} \\
\midrule
7 & Mean   & ASP & 1   
& 0.652 & 0.745 & 0.785 & 0.442 & 0.708 & 0.218 & 0.664 & 0.871 & 0.557 & 0.863
\\
8 & Mean   & ASP & 5   
& 0.656 & 0.733 & \textbf{0.795} & \textbf{0.417} & \textbf{0.717} & 0.218 & 0.654 & 0.893 & \textbf{0.583} & \textbf{0.820}
\\
9 & Mean   & ASP & 64  
& 0.679 & 0.754 & 0.790 & 0.427 & 0.695 & 0.226 & 0.670 & 0.857 & 0.562 & 0.874
\\
10 & Mean   & ASP & 128 
& 0.653 & 0.744 & 0.792 & 0.422 & 0.710 & 0.218 & \textbf{0.673} & \textbf{0.852} &  0.580 & 0.828
\\
%\cmidrule{1-14}
%\multicolumn{4}{c||}{Average} & 0.660 & 0.744 & 0.791 & 0.427 & 0.708 & 0.220 & 0.665 & 0.868 & 0.571 & 0.846 \\
\midrule

11 & 12   & ASP & 1   & 0.652 & 0.801 & 0.795 & 0.420 & 0.682 & 0.228 & 0.591 & 1.014 & 0.574 & 0.834 \\
12 & 12   & ASP & 5   & 0.661 & 0.745 & 0.795 & 0.409 & 0.696 & 0.219 & 0.607 & 0.995 & 0.574 & 0.838 \\
13 & 12   & ASP & 64  & 0.669 & 0.764 & 0.789 & 0.422 & 0.666 & 0.239 & 0.647 & 0.905 & 0.565 & 0.860 \\
14 & 12   & ASP & 128 & 0.660 & 0.756 & \textbf{0.797} & \textbf{0.405} & 0.666 & 0.236 & 0.644 & 0.908 & 0.569 & 0.852 \\
%\cmidrule{1-14}
%\multicolumn{4}{c||}{Average} & 0.660 & 0.767 & 0.794 & 0.414 & 0.677 & 0.231 & 0.622 & 0.956 & 0.570 & 0.846 \\
%\midrule
%\multicolumn{14}{c}{\textbf{ASP OVER TIME AND LAYER}} \\
%\midrule
%\multicolumn{4}{c||}{Best Models} & 0.68 & \textbf{0.72} & 0.79 & 0.44 & 0.71 & \textbf{0.21} & 0.66 & 0.90 & 0.55 & 0.86 \\
\bottomrule
\end{tabular}} }
\label{tab:experimental_results}
\end{table*}

\subsection{Attentive Statistics Pooling (ASP)}\label{sec:stat_pooling}
%As a layer-wise and time-wise pooling mechanism, we adopted channel- and context-dependent attentive statistics pooling, introduced in the ECAPA-TDNN architecture for speaker verification~\cite{ecapa_tdnn}.
For both layer- and time-wise pooling, we adopt channel- and context-dependent attentive statistics pooling
%~\cite{okabe18_interspeech} 
%from the ECAPA-TDNN architecture~
\cite{ecapa_tdnn} provided by the SpeechBrain toolkit~\cite{speechbrain}.
An overview of the process is provided in component (2) of Figure \ref{fig_overview_architecture}. 
Input features are first projected into a low dimensional space via a Time-Delay Neural Network~\cite{tdnn} (TDNN) block.
%, consisting of a point-wise 1D convolution, ReLU, and batch normalization. 
After the TDNN block, the Tanh activation function is applied. 
%The resulting dimensions can be interpreted as independent attention heads, each selectively encoding different aspects of temporal dynamics. 
To capture channel-specific attention, the feature representations are projected back to the original feature dimensionality through a 1D convolution with a kernel size of 1, transforming them into attention scores.
%Subsequently, a softmax normalization is applied along the time dimension independently for each channel, resulting in importance weights across time.
Subsequently, the softmax function is applied along the time dimension independently for each channel, resulting in temporal importance weights.
Using these normalized attention weights, the pooling mechanism computes the weighted mean and weighted standard deviation of the input features over time. 
%Our attentive statistics pooling implementation is adapted from the ECAPA-TDNN model provided by the SpeechBrain toolkit~\cite{speechbrain}.
The final output of the attentive statistics pooling is given by
\begin{equation*}
\mathbf{z} = concat(\mu^{\text{att}}, \sigma^{\text{att}}) \in \mathbb{R}^{2d},
\end{equation*}
where $\mu^{\text{att}}$ and $\sigma^{\text{att}}$ denote the attention-weighted mean and standard deviation over the time dimension, respectively, and $d$ is the dimensionality of the input feature vectors. 

\vspace{-0.6em}
\subsubsection{Layer-wise Representations}
To investigate whether weighting information across all layers of the W2V2-based feature extractor provides richer and more task-relevant representations than relying on time-wise representations, we adopt a layer-wise ASP mechanism. 
After feature extraction using W2V2, mean pooling is applied over the time dimension for each individual transformer layer, resulting in 24 separate representations. These representations are concatenated to form a feature matrix of dimension $\mathbb{R}^{24 \times 1024}$.
%, where each row corresponds to a distinct transformer layer.
ASP
%, as outlined in Section \ref{sec:stat_pooling}, 
is applied to produce a representation that captures weighted statistical information across all layers. 

\subsubsection{Time-wise Representations}
To examine the temporal characteristics of the W2V2-based features, 
%in a manner that is directly comparable to the layer-wise representations group, 
we compute the mean across all transformer layers of the W2V2 feature extractor. 
The resulting features are weighted and aggregated over the temporal dimension using the ASP mechanism described in Section~\ref{sec:stat_pooling}, allowing the model to consider the differing relevance of each time frame.
In addition to the mean-layer representation, we apply ASP to representations from a single transformer layer to compare their performance with the averaged representation. 
%For illustration, we focus on features from layer 12. 
We selected layer 12 for the comparison, motivated by prior work~\cite{nguyen_asr_slt2024, layer_problem_pathological_speech}, which has demonstrated that intermediate layers are particularly informative for processing pathological speech, including the prediction of intelligibility and severity.

\section{Experiments}
% conducted
All experiments were formulated as regression tasks targeting the speech descriptors 
\textit{intelligibility}, \textit{harsh voice}, \textit{monoloudness}, \textit{inappropriate silences} and \textit{imprecise consonants}. 
%In addition to the attention methods described in the previous section, traditional 
Besides the ASP methods described earlier, mean pooling without attention is used as a baseline pooling mechanism on the mean layer representations (Exp.\ 1) and layer 12 representations (Exp.\ 2). 
All models were trained using the Adam~\cite{adam_optimizer} optimizer with \(\beta_1 = 0.9\) and \(\beta_2 = 0.999\), % and early stopping implemented based on development loss and a patience of 15 epochs. The learning rate was fixed at $10^{-5}$ and the batch size was uniformly set to 32.
early stopping after 15 epochs, a fixed learning rate of $10^{-5}$, and a batch size of 32. 
For each model configuration, attention heads \( a_h \in \{1, 5, 64, 128\} \) were evaluated. 
Model performance was assessed using mean squared error (MSE) and the Pearson correlation coefficient (PCC). To compare model and group performance, pairwise t-tests on the MSE were conducted to assess statistical significance between the models at a 5\% significance level. For comparisons of ASP with the baseline, t-tests were conducted between the best-performing ASP model and the baseline for each speech descriptor. For the comparison of ASP over time and ASP over layer,
%the ASP groups, 
significance testing was performed between both ASP experiments (i.e., Exp.\ 3-6 vs. Exp.\ 7-10) per speech descriptor.

\section{Results and Discussion}

\begin{figure*}[tb]
\centerline{\includegraphics[width=0.95\linewidth]{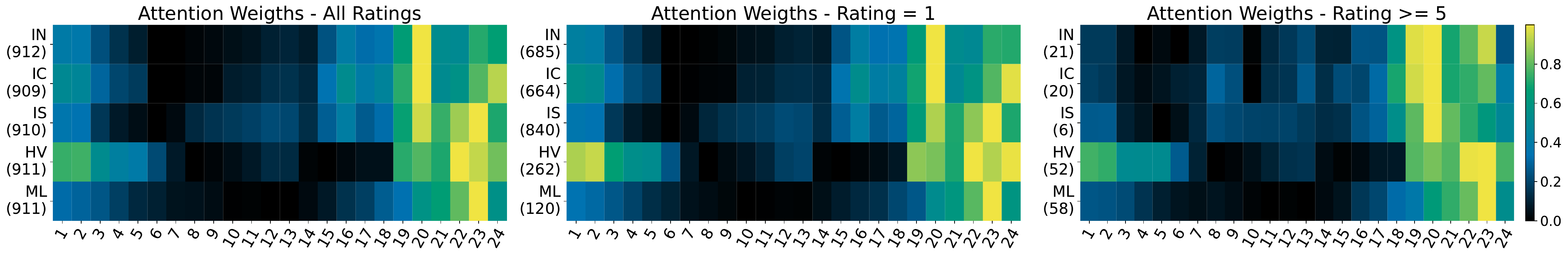}}
\caption{Attention weight distributions derived from the best-performing layer-wise ASP models, chosen with respect to the PCC metric and analyzed by ground-truth rating levels (1=typical, 7=severe). 
Weights were averaged across feature dimension and subsequently across all selected samples. To ensure comparability, the values were scaled to a range of $\left[0,1\right]$. X-axis represents the layers of the model. Y-axis format: Descriptor abbreviation (number of samples). Abbreviations: IN = Intelligibility, IC = Imprecise consonants, IS = Inappropriate silences, HV = Harsh voice, ML = Monoloudness.
}
\label{fig_layerwise_att_weights}
\end{figure*}

Table \ref{tab:experimental_results}
provides an overview of the experimental results comparing time- and layer-wise ASP mechanisms across various attention head configurations. Overall, the comparison of ASP models with the baseline (Exp.\ 1-2) demonstrates that the best performance is consistently achieved using an ASP method, with MSE values significantly lower than the baseline, underscoring the benefit of incorporating attention mechanisms into the pooling process.

Models with five attention heads most frequently achieve the best results, indicating that a moderate number of heads is in most cases sufficient to capture relevant information in the selected speech descriptors. For the descriptors \textit{intelligibility} and \textit{inappropriate silences}, models with a higher number of attention heads achieved slightly better performance in terms of MSE.
%, but the improvement did not exceed 0.01 compared to the best-performing model within the ASP group.
Only for \textit{harsh voice}, using 128 attention heads resulted in the most substantial improvement, with the MSE decreasing by 0.041 relative to the configuration with five heads. Across the other configurations of \textit{harsh voice} with 1 and 64 heads, only deviations of about 0.01 in MSE were observed, suggesting that the poor performance with five attention heads may not be caused by their number. Overall, the number of attention heads has little effect, indicating that the choice between time-wise and layer-wise representations is more crucial for regressing the selected speech descriptors. According to PCC and MSE metrics, all speech descriptors show an advantage with one of the ASP groups, except for inappropriate silences, which perform similarly under both approaches.

Notably, the descriptor \textit{harsh voice} shows the highest MSE values across all experiments ($0.852 \leq x \leq 1.059$), but given the 1-to-7 label range, the ratings remain largely reasonable. This is supported by high PCC values, reflecting a strong linear correlation with the ground truth despite the increased MSE.

\subsection{Layer-wise Representations}\label{sec:layer_results}
An analysis of the layer-wise ASP experiment results (Exp.\ 3-6) 
% in Table~\ref{tab:experimental_results}
indicates that regression performance for the speech descriptor \textit{intelligibility} benefits from the use of layer-wise representations. The layer-wise ASP models group for intelligibility achieved a significantly lower MSE compared to the time-wise ASP models group (Exp.\ 7-10). One possible explanation 
% for this result 
is that \textit{intelligibility} represents a holistic metric, which in human perception relates to the comprehensibility of an entire utterance rather than the evaluation of individual speech segments. Thus, representations averaged over the complete utterance appear to be sufficient, and the model benefits more from the integration of information across different transformer layers than from the explicit modeling of temporal aspects.

In Figure \ref{fig_layerwise_att_weights}, the averaged attention weights across the layer dimension, scaled to a range of $\left[0,1\right]$, are depicted with their corresponding ground-truth ratings. 
Attention weight distributions across all ratings show that most weight is placed on the first three and last six layers, while intermediate layers receive little attention, except for \textit{inappropriate silences}, which also emphasizes intermediate layers to some degree. \textit{Harsh voice}, a criterion that primarily reflects phonatory characteristics, shows attention weights concentrated in the first layers more prominently than the other descriptors.
% The visualization of attention weighting across all ratings shows that in most cases higher attention weighting is assigned to the first three and last six layers, while intermediate layers receive less weighting, except for \textit{inappropriate silences}, which also emphasizes intermediate layers. 
Overall, this pattern may be attributed to the W2V2 model's fine-tuning on ASR, where later layers may primarily be responsible for linguistic and semantic processing, while earlier layers may focus on encoding acoustic features, consistent with the findings of Pasad et al.\ \cite{layerwise_analysis_asru}. 
Attention distribution further varies with severity: for most descriptors, unaffected speech (rating 1) emphasizes initial and final layers, while for severely affected speech (rating $\geq 5$), attention in most cases shifts to intermediate and later layers.
Overall, the layer-wise ASP experiments demonstrate that no single layer within the W2V2 feature extractor is exclusively preferred through the learned attention weights. Instead, the model appears to leverage information distributed across multiple layers, often drawing from the earlier and later layers. 
%Layer-wise attention consistently outperforms time-wise attention for three of the five speech descriptors, highlighting the value of integrating representations across the depth of the W2V2 feature extractor.

\subsection{Time-wise Representations}
\label{ssec:exp_timewise}
Table~\ref{tab:experimental_results} shows that regression performance for \textit{imprecise consonants}, \textit{harsh voice}, and \textit{monoloudness} benefits from time-wise ASP. For \textit{imprecise consonants} and \textit{harsh voice}, the time-wise ASP models group (Exp.\ 7–10) achieves a significant improvement over layer-wise ASP models (Exp.\ 3–6) in terms of MSE.
Although \textit{monoloudness} achieves the highest PCC and lowest MSE with a time-wise ASP model, no statistically significant reduction in MSE compared to the layer-wise ASP models group can be observed.
Taken together, these findings could potentially be explained by the temporal characteristics of the descriptors: \textit{imprecise consonants} require observation of multiple consonant productions over time, \textit{harsh voice} benefits from capturing sustained phonatory patterns that occur over extended vocalizations, and \textit{monoloudness} depends on variations in loudness across longer speech segments.
Models that preserve temporal resolution, such as time-wise ASP, may therefore be better suited to capture these temporally distributed acoustic cues.

In addition to the comparison between time-wise and layer-wise ASP experiments, we further examine the role of single-layer representations by comparing the mean-layer ASP models (Exp.\ 7–10) with the 12th-layer ASP models (Exp.\ 11–14). The baseline results (Exp.\ 1, 2) indicate that using only layer 12 yields in most cases performance similar to the mean of all layers ($\pm$0.01), except for \textit{harsh voice}, where the 12th-layer representation performs considerably worse. As already discussed in Section~\ref{sec:layer_results}, this may reflect the hierarchical separation of information across layers, suggesting that earlier layers tend to capture acoustic and phonetic information, which might be largely absent in layer 12.
These baseline results are only partially transferable to the time-wise ASP setting. For most speech descriptors, the 12th-layer representation performs similarly or worse than the mean-layer representation, with the exception of \textit{imprecise consonants}, where the 12th-layer representations provide a slight improvement (-0.012 MSE, +0.002 PCC).
The \textit{intelligibility} results confirm the layer- vs.\ time-wise findings, showing that, across all attention head configurations, 12th-layer ASP models perform worse in terms of MSE than their layer-mean counterparts.
%suggesting that intelligibility ratings rely on features distributed across the representational hierarchy rather than those captured exclusively in the 12th layer. 
A similar tendency is observed for the \textit{harsh voice} models. This mirrors the baseline pattern and indicates that inter-layer information plays an important role in predicting \textit{harsh voice}. Nevertheless, when comparing time-wise with layer-wise ASP, time-wise representations remain clearly superior for predicting the \textit{harsh voice} rating, highlighting that both temporal resolution and multi-layer integration contribute distinct and complementary benefits to modeling this speech descriptor.

\section{Conclusion}
In this work, we investigated the impact of layer-wise and time-wise ASP mechanisms using features extracted from a W2V2-based model.
Our results indicate that \textit{intelligibility} is more effectively modeled with layer-wise representations, whereas \textit{imprecise consonants}, \textit{harsh voice} and \textit{monoloudness} are better captured with time-wise modeling. For \textit{inappropriate silences}, no clear advantage could be observed for either group. Furthermore, an analysis of attention weights reveals that layer-wise attention distributions vary with the severity of the speech descriptors, suggesting that different severity levels activate distinct layers.
A comparison between single-layer and multi-layer time-wise ASP experiments reveals that inter-layer information still plays an important role for certain descriptors, even though time-wise ASP provides more informative representations than layer-wise ASP for these descriptors. This observation motivates future work on combining layer-wise and time-wise ASP. 
% Initial experiments suggest that a combined approach can achieve performance comparable to the best models using separate ASP mechanisms, although further investigation is needed to fully explore its potential.

\pagebreak

\bibliographystyle{IEEEbib}
\footnotesize{
\bibliography{refs}
}

\end{document}